\documentclass{article}
\usepackage[square,sort,comma,numbers]{natbib}

\usepackage[preprint]{neurips_2019}

\usepackage[utf8]{inputenc} 
\usepackage[T1]{fontenc}    
\usepackage{hyperref}       
\usepackage{url}            
\usepackage{booktabs}       
\usepackage{amsfonts}       
\usepackage{nicefrac}       
\usepackage{microtype}      

\usepackage{amsmath, amssymb, amsfonts, dsfont,amsthm}
\usepackage{color}
\usepackage{bm}
\usepackage{subcaption}
\usepackage{caption}
\usepackage{enumitem}

\DeclareMathOperator{\Pro}{Pr}
\DeclareMathOperator{\Exp}{\mathds{E}}
\DeclareMathOperator{\Mult}{Mult}
\DeclareMathOperator{\Dir}{Dir}
\DeclareMathOperator*{\plim}{plim}
\DeclareMathOperator{\Ber}{Bernoulli}
\DeclareMathOperator{\Unif}{U}
\DeclareMathOperator{\dd}{d}
\newtheorem{prop}{Proposition}
\newenvironment{prf}{%
	\begin{proof}%
	}{%
	\end{proof}%
}

\usepackage{tikz}
\usetikzlibrary{fit,positioning}
\usepackage[title]{appendix}

\begin{document}
	\title{Latent Distribution Assumption for\\Unbiased and Consistent Consensus Modelling}

	\author{
		Valentina Fedorova \quad Gleb Gusev \quad Pavel Serdyukov\\
		Yandex\\
		16 Leo Tolstoy St.\\
		Moscow\\
		Russia\\
		119021\\
		\texttt{\{valya17,gleb57,pavser\}@yandex-team.ru}
	}
	
	\maketitle

	\begin{abstract}
		We study the problem of aggregation noisy labels. 
		Usually, it is solved by proposing a stochastic model for the process of generating noisy labels and then estimating the model parameters using the observed noisy labels.
		A traditional assumption underlying previously introduced generative models is that each object has one \emph{latent true label}.
		In contrast, we introduce a novel \emph{latent distribution assumption}, implying that a unique true label for an object might not exist, but rather each object might have a specific distribution generating a \emph{latent subjective label} each time the object is observed.
		Our experiments showed that the novel assumption is more suitable for difficult tasks, when there is an ambiguity in choosing a ``true'' label for certain objects.
		
	\end{abstract}
	
	\section{Introduction}
	
	Crowdsourcing marketplaces, such as Amazon Mechanical Turk\footnote{https://www.mturk.com}, make it possible to label large data sets in a shorter time as well as at a lower cost comparing to that needed for a limited number of experts. However, as workers at the marketplaces are non-professional and vary in levels of expertise, such labels are much noisier than those obtained from experts.
	In order to reduce the noisiness, typically, each object is labelled by several workers, and then these labels are further aggregated in a certain way to infer a more reliable \emph{consensus label} for the object.
	Most advanced consensus models \citep[e.g.,][]{dawid/skene:1979,whitehill/etal:2009,zhou/etal:2015} address different aspects of uncertainty in the process of generating noisy labels.
	
	A \emph{traditional setting} used in those and other previous studies is based on the \emph{latent label assumption (L-assumption)}, implying that
	each object has a unique \emph{latent true label}, and, when a worker observes the object, this latent true label is corrupted with regard to a chosen stochastic model into an \emph{observed noisy label}. As a consequence, consensus models designed under this assumption explain any disagreements among observed noisy labels of an object by the mistakes made by some of the labellers.
	However,
	this may not explain a certain kind of disagreements among labels produced by experts, which is typical for some object domains.
	E.g., when assessing relevance of documents to search queries, even well trained experts may disagree about the true label for certain objects \citep[see e.g.,][]{voorhees:2000}. 
	This is equivalent to saying that a unique true label of an object does not exist, but rather each object has its specific distribution over possible \emph{subjective labels}, which is induced by a distribution of personal preferences over different aspects of the task.
	This type of uncertainty lies beyond the traditional L-assumption, and this paper introduces a novel approach based on the \emph{latent distribution assumption (D-assumption)} to deal with this problem.
		
	The novel D-assumption suggests the following generative process: Each object has its specific (\emph{latent}) distribution over subjective values of label. Each time a worker observes the object, a latent subjective label is sampled from the object's distribution. Then, this subjective label is corrupted according to a stochastic model, and an observed label is revealed. It is crucial that the object's subjective label introduced in the process is generated each time a worker observes the object. 
	This distinguishes the latent distribution approach from the previous consensus models.
	
	Besides, to the best of our knowledge, this is the first paper that looks at the output probabilities of consensus labels from a statistical point of view. In particular, we notice that posterior probabilities of labels obtained under the traditional L-assumption are poor estimates of the underlying true probabilities. In contrast to this, we show, both analytically and experimentally, that the probabilities obtained under the D-assumption may serve as accurate statistical estimates of the underlying true probabilities. 
	Thus, the latent distribution for subjective labels of an object estimated via our framework can be further used for the problems of label distribution learning considered in \citep{geng:2016}.
	
	\noindent\textbf{Background. }
	A line of previous studies on consensus modelling \citep[e.g.,][]{dawid/skene:1979,whitehill/etal:2009} explicitly assumed that each object has a single ``true'' label. Other works \citep[e.g.,][]{zhou/etal:2012,zhou/etal:2014,zhou/etal:2015} including  Bayesian approaches \citep[e.g.,][]{bachrach/etal:2012,kim/ghahramani:2012,venanzi/etal:2014},
	made the L-assumption implicitly: though it is assumed that each object is associated with a probabilistic label, it is not allowed that a latent label of an object takes multiple different values. In other words, posterior probabilities for latent labels, inferred under the traditional L-assumption, is a measure of confidence in each value being a unique latent true label.
	Our work is orthogonal to those studies as it considers a different assumption for the process of generating noisy labels and leads to essentially different and ``better calibrated'' output probabilities. \citep{nguyen/etal:2016} suggested a consensus model for subjective tasks, our work is more fundamental in the sense that we describe a general framework to consensus modelling in such tasks and theoretically study properties of their outputs.

	\section{Latent distribution assumption}\label{sec:latent_distribution_concept}
	
	Let \(Z_j\) be a random variable whose value is the latent label for an object \(j\), and \(Y^w_j\) be a random variable whose value is the observed noisy label assigned by a worker \(w\) to an object \(j\). 
	
	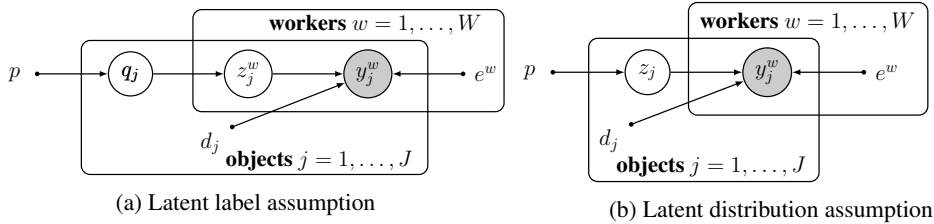
\begin{figure}[t]
		\centering
		\begin{subfigure}{.49\textwidth}
			\centering
			\resizebox {1.3\textwidth} {!} {
				\begin{tikzpicture}
				\tikzstyle{main}=[circle, thick, minimum size = 10 mm, font = \Large, node distance = 16mm]
				\node[main] (p) {\(p\) }; \fill[black] (p.east) circle(1.5pt);
				\node[main, draw = black] (Qj) [right=of p] {\(q_j\)};
				\node[main] (Qj) [right=of p] {\(q_j\)}; 
				\node[main, draw = black] (zjw) [right=of Qj] {\(z^w_j\)};
				\node[main, draw = black, fill = black!20] (yjw) [right=of zjw] {\(y^w_j\)};
				\node[main,] (e) [right=of yjw] {\(e^{w}\)}; \fill[black] (e.west) circle(1.5pt);
				\node[main, node distance = 5mm, font = \Large, label= left: {\Large{\(d_j\)} }] (d) [below=of zjw] {}; \fill[black] (d.north west) circle(1.5pt);

				\node[rectangle,  inner sep=0mm, fit= (e)(yjw)(zjw), label= above left: {\Large{\textbf{ workers \(w = 1,\ldots,W\)}}}, xshift = 33mm ,yshift=2mm] {};
				\node[rectangle, rounded corners = 2mm,  inner xsep=3mm, inner ysep=6mm, draw = black, thick, fit= (e)(yjw)(zjw), xshift = -4mm ,yshift=3mm] {};
				
				\node[rectangle,   fit=(yjw)(zjw)(d)(Qj), label= below: {\Large{{\textbf{ objects \(j = 1,\ldots,J\)}}}}, xshift = 15mm ,yshift=6mm] {};
				\node[rectangle, rounded corners = 2mm,  inner xsep=7mm, inner ysep=2mm, draw = black,thick, fit=(yjw)(zjw)(d)(Qj), xshift = 1mm ,yshift=0mm] {};

				\foreach \from/\to in {e/yjw,zjw/yjw,p/Qj,Qj/zjw,d.north west/yjw}
				\draw [->, thick, draw = black, -latex] (\from) -- (\to);
				\end{tikzpicture}
			}
			\caption{Latent label assumption}
			\label{fig:graphical-models-traditional}
		\end{subfigure}%
		\begin{subfigure}{.51\textwidth}
			\centering
			\resizebox{1.1\textwidth}{!}{
			\begin{tikzpicture}
			\tikzstyle{main}=[circle, thick, minimum size = 10 mm, font = \Large, node distance = 16mm]
			\node[main] (p)  {\(p\) }; \fill[black] (p.east) circle(1.5pt);
			\node[main, draw = black] (zj) [right=of p] {\(z_j\)};
			\node[main, node distance = 5mm, label= left: {\Large{\(d_j\)} }] (d) [below=of zj] {}; \fill[black] (d.north west) circle(1.5pt);
			\node[main, draw = black, fill = black!20] (yjw) [right=of zj] {\(y^w_j\)};
			\node[main] (e) [right=of yjw] {\(e^{w}\)}; \fill[black] (e.west) circle(1.5pt);
			
			\node[rectangle,  inner sep=0mm, fit= (e)(yjw)(zj), label= above left: \Large{{\textbf{ workers \(w = 1,\ldots,W\)}}}, xshift = 40mm ,yshift=2mm] {};
			\node[rectangle, rounded corners = 2mm,  inner xsep=8mm, inner ysep=7mm, draw = black,thick, fit= (e)(yjw), xshift = -4mm ,yshift=3mm] {};
			
			\node[rectangle,   fit=(yjw)(zj)(d), label= below: {\Large{{\textbf{ objects \(j = 1,\ldots,J\)}}}}, xshift = 0mm ,yshift=4mm] {};
			\node[rectangle, rounded corners = 2mm,  inner xsep=7mm, inner ysep=3mm, draw = black,thick, fit=(yjw)(zj)(d), xshift = -1mm ,yshift=-1mm] {};

			\foreach \from/\to in {e/yjw,zj/yjw,p/zj,d.north west/yjw}
			\draw [->, thick, draw = black, -latex] (\from) -- (\to);
			\end{tikzpicture}
		}
			\caption{Latent distribution assumption}
			\label{fig:graphical-models-new}
		\end{subfigure}
		\caption{Graphical structures for two generative models based on different assumptions.}
		\label{fig:test}
		\vspace{-5mm}
	\end{figure}
	To the best of our knowledge, all existing consensus models do not go beyond the general generative model shown in Figure~\ref{fig:graphical-models-traditional}. This class of models, including \citep{dawid/skene:1979,whitehill/etal:2009,zhou/etal:2012} and others, which we call \emph{the LA-model}, represents the following generative process:
	\begin{enumerate}
		\item A unique latent true label \(z_j\) for object \(j\) is drawn from a probability distribution \(P_{Z}\) with a parameter \(p\) (note that \(P_{Z}\) is the same for all objects). For multiclass labels, \(P_Z\) is usually a multinomial distribution \(\Mult(p)\), where \(p\) is a vector of prior probabilities.
		\item Given the latent true label \(z_j\) for an object \(j\), an observed noisy label \(y_j^w\) from a worker \(w\) for this object is sampled from the conditional probability distribution \(\Pro(Y_j^w \mid Z_j =z_j)\)\footnote{As usual, capital letters are used for random variables and the corresponding lower case letters are for particular values of the random variables.} with two potential parameters\footnote{``Potential parameters'' means that some of the parameters may be absent.} \(e^w\) and \(d_j\). Intuitively, \(e^w\) represents a level of expertise for the worker \(w\), and \(d_j\) corresponds to a level of difficulty for the object \(j\).
	\end{enumerate}
	Note that most previous studies are focused on the second step. They proposed different ways to model the corruption of the true label and, for example, additionally employed latent communities of workers  \citep{venanzi/etal:2014} or features of the object and the worker \citep{ruvolo/etal:2013}. However, this does not put these models beyond the general two-steps scheme described above.
	
	In this paper, by using the novel D-assumption we change the first step and introduce the novel class of models, which we call the \emph{DA-model}. At this point, for each object \(j\), we define a random variable \(Q_j\) whose value \(q_j\) is a parameter of the distribution of subjective labels for this object. The graphical model is shown in Figure~\ref{fig:graphical-models-new} and represents the following generative process: %
	\begin{enumerate}
		\item For each object \(j\), a parameter \(q_j\) is drawn form a probability distribution \(P_Q\) parameterised by \(p\) (again, note that \(P_Q\) is the same for all objects). 
		
		\item When a worker \(w\) observes an object \(j\):
		\begin{enumerate}
			\item A subjective label \(z^w_j\) is sampled from the object's distribution with the parameter \(q_j\).
			\item A noisy label \(y^w_j\) is generated from the conditional probability distribution \(\Pro(Y_j^w \mid Z_j = z^w_j)\) with two potential parameters \(e^w\) and \(d_j\).
		\end{enumerate}
	\end{enumerate}
	Observe that the key difference between Figures~\ref{fig:graphical-models-traditional} and \ref{fig:graphical-models-new} is in the workers' plate: for the DA-model, the workers' plate is extended over the latent subjective labels~\(z_j^w\). Next we show that this difference constitutes a conceptual change for the inference of consensus labels.
	
	Consider a set of noisy labels \(\{y^w_j\}\) for objects \(j \in J\) from workers \(w \in W\).
	Traditionally, consensus labels are inferred by maximising the log-likelihood of the observed noisy labels. For the L-assumption, the consensus output for object \(j\) is probability \(\Pro(Z_j = z_j)\) for each possible value \(z_j\), reflecting our confidence that the value \(z_j\) is the unique true label for object \(j\). In this setting, the log-likelihood of the observed data is
	\begin{align}\label{eq:target_ll}
	\begin{small}
	\sum_{j\in J}\log\bigg[\sum_{z_j\in \mathcal{Y}}\Pro(z_j\mid p) \prod_{w \in W_j} \Pro(y_j^w \mid z_j)\bigg],
	\end{small}
	\end{align}
	where the second sum is taken over the set \(\mathcal{Y}\) of possible label values, \(W_j\) is the set of indices of workers' who labelled \(j\)-th object, and notations of the form \(\Pro(x)\) are used instead of \(\Pro(X=x)\) for short, and for the same reason we omitted possible parameters \(e^w\) and \(d_j\) in \(\Pro(y_j^w \mid z_j)\). Whereas, under the D-assumption, the consensus output for object \(j\) is the distribution of subjective labels \(z^w_j\), parameterized by the object's parameter \(q_j\). And then the log-likelihood is defined as
	\begin{align*}
	\begin{small}
	\sum_{j\in J}\log\int \Pro(q_j \mid p)\prod_{w \in W_j}\bigg[\sum_{z^w_j\in \mathcal{Y}} \Pro(z^w_j\mid q_j)\Pro(y_j^w \mid z_j^w)\bigg] \dd q_j.
	\end{small}
	\end{align*}
	To make the discussion of DA-models easier, in the rest of this paper we assume that parameter \(p\) of the DA-model is a deterministic distribution assigning to each object a chosen beforehand (but unknown) distribution of the object's subjective labels. Then, we can remove the parameter \(p\) in Figure~\ref{fig:graphical-models-new} and assume that \(q_j\) are additional parameters of the model, and the expression for the log-likelihood of the observed labels becomes the following:
	\begin{align}\label{eq:target_ld}
	\begin{small}
	\sum_{j\in J}\sum_{w \in W_j}\log\bigg[\sum_{z^w_j\in \mathcal{Y}} \Pro(z^w_j\mid q_j)\Pro(y_j^w \mid z_j^w)\bigg].
	\end{small}
	\end{align}
	
	\noindent\textbf{Remark.} Two described approaches resemble, to some extent, two approaches to topic modelling -- the mixture of unigrams model and the latent Dirichlet allocation model \citep[see, e.g.,][]{blei/etal:2003}. The first model assumes that each document is associated with a single topic, which specifies the distribution of words in this document; for the second model, each document may be associated with multiple topics, and words in this document are generated from a mixture of the distributions for the topics. By replacing topics for latent labels and words for noisy labels, we get the two approaches discussed in this section.
	Note, however, that for our setting, it is necessary that the domains of latent and observed labels are the same. So that latent ``topics'' are in one-to-one correspondence with the label values. Otherwise, we may not interpret the latent distribution.

	Finally, let us motivate the novel approach by the following numerical example. Consider 210 workers with highly confident estimates of their expertise inferred from their labels for many tasks they completed. Let each worker assigns the correct label with probability \(a:=0.8\) (independent of the value of that correct label). Consider a binary classification task with uninformative prior over the two labels \(\{0,1\}\). Let each of the workers provides one noisy label for a new object, and it turns out that 110 of the labels are 1s and 100 are 0s. Under the L-assumption, the posterior probability \(p_1\) of the true label being 1 is \(p_1 \approx a^{10}/(a^{10}+(1-a)^{10})> 1-2^{-20}\), i.e. the correct label is practically 1 with probability 1, and we observe a very unreliable event that 100 out of 210 workers made a mistake. On the contrast, under the D-assumption,
	according to \eqref{eq:target_ld}, the probability \(p_1\) should maximize $110 \log(0.8 p_1+0.2 (1-p_1)) + 100 \log (0.2 p_1+ 0.8 (1-p_1))$, and we infer that most of the workers provided true (though subjective) labels and
	for this object the latent Bernoulli distribution has the parameter \(p_1 \approx 0.54\), which seems much more realistic.

	\section{Consensus models under L- and D-assumptions}\label{sec:new_models}
	
	A distinguishing feature of the DA-model, shown in Figure~\ref{fig:graphical-models-new}, is that latent subjective labels for an object are sampled each time the object is observed. Thus, given a certain LA-model, the corresponding DA-one is defined as follows:
	Firstly, the distribution \(P_Q\) in the DA-model is defined over the domain of the parameter \(p\) in the LA-model. E.g., for the examples of models  below, \(p\) is the parameter of the multinomial distribution \(P_Z\), and \(P_Q\) is defined to be a Dirichlet distribution.
	Secondly, The conditional distribution \(\Pro(Y_j^w \mid Z_j = z^w_j)\) in the DA-model is the conditional distribution of noisy labels defined in the LA-model, with the subjective label value \(z^w_j\) for the random variable \(Z_j\).
	
	The rest of this section describes three established models\footnote{For all the models, let labels take values from \(\mathcal{Y} := \{1, \ldots, K\}\).} as special cases of the traditional LA-model, and their novel counterparts as special cases of the novel DA-model. Later, in Section~\ref{sec:empirical}, we will empirically compare the performance of all these models.
	
	\noindent\textbf{3.1\quad Dawid and Skene model}\\
	Consider a special case of the LA-model with parameters \(p\) and \(\{e^w\}\) defined as follows: \(p := (p_1, \ldots, p_K)\) is the vector of prior probabilities for label values; the parameter \(e^w\) is the confusion  matrix of size \(K\times K\), and \(\Pro(Y^w_j = y^w_j \mid Z_j = z_j ) := e^w(z_j, y^w_j)\). The model was proposed in \citep{dawid/skene:1979} and we will refer to it as \emph{LA DS}.
	
	The corresponding special case of the DA-model is the following: (1) for each object \(j\), a vector \(q_j = (q_{j,1},\ldots,q_{j,K})\) is drawn from a Dirichlet distribution \(P_Q := \Dir(p)\), this vector \(q_j\) is the parameter of the multinomial distribution of subjective labels for this object; (2) when a worker \(w\) observes an object \(j\), first, a subjective label \(z^w_j\) is drawn from the multinomial distribution \(\Mult(q_j)\), and then, a noisy label \(y^w_j\) is drawn from the multinomial distribution \(\Pro(Y_j^w \mid Z_j = z^w_j) := \Mult(e^w(z^w_j, \cdot))\), where \(e^w(k, \cdot)\)  stands for \(k\)-th row of the confusion matrix \(e^w\). It will be denoted as \emph{DA DS}.
	
	\noindent\textbf{3.2\quad GLAD}\\
	Consider the following special case of the LA-model with parameters \(\{e^w\}\), and \(\{d_j\}\): parameter \(p \) is absent meaning that true labels \(z_j\) are assumed to be deterministic; the parameter \(e^w\) is a scalar value representing the level of expertise for the worker \(w\); and the parameter \(d_j\) is a scalar value representing the level of difficulty for the object \(j\). Let \(a(w,j):=\frac{1}{1 + \exp{(-e^w d_j)}}\) and define
	\begin{multline*}
	\Pro(Y_j^w = y\mid Z_j = z_j) := P_{\rm GLAD}(Y^w_j\mid Z_j = z_j) =\left\{
	\begin{array}{ll}
	\!\!\!\!a(w,j) , &\!\!\!\! {\rm for~}y = z_j; \\
	\!\!\!\!\frac{1 - a(w,j)}{K - 1}, &\!\!\!\! {\rm for~}y \in \{1,\ldots,K\} \setminus z_j. \\
	\end{array}
	\right.
	\end{multline*}
	This model is called GLAD\footnote{GLAD is a shortcut for Generative model of Labels, Abilities, and Difficulties.} and was described in \citep{whitehill/etal:2009}, it will be denoted as \emph{LA GLAD}.
	
	The corresponding DA-model is the following: (1) each object \(j\) has a deterministic parameter \(q_j\in [0,1]\); (2) for a worker \(w\) and an object \(j\), a latent subjective label \(z^w_j\) is sampled form the multinomial distribution \(\Mult(q_j)\), and then, a noisy label \(y^w_j\) is generated from \(\Pro(Y_j^w \mid Z_j = z^w_j) := P_{\rm GLAD}(Y^w_j \mid Z_j = z^w_j)\). We will refer to it as \emph{DA  GLAD}.
	
	\noindent\textbf{3.3\quad Minimax entropy principle}\\
	Consider a model described in \citep[Section 2]{zhou/etal:2015} for multiclass labels; this model will be referred to as \emph{LA MME}. The model was derived using a minimax entropy principle, and it is a special case of the LA-model with \(\{e^w\}\) and \(\{d_j\}\) defined as matrices of size \(K\times K\). Using these parameters, for each worker \(w\) and each object \(j\), define \(\Pro(Y^w_j = y^w_j\mid Z_j = z_j ) := P_{\rm MME}(Y^w_j \mid Z_j = z_j)\propto \exp{\left(e^w(z, y) + d_j(z, y)\right)}\).
	
	The DA-counterpart is the following: (1) each object \(j\) has a deterministic parameter \(q_j\in [0,1]\); (2) for a worker \(w\) and an object \(j\), a latent subjective label \(z^w_j\) is sampled from the multinomial distribution \(\Mult(q_j)\), and then, a noisy label \(y^w_j\) is generated from \(\Pro(Y_j^w \mid Z_j = z^w_j) :=P_{\rm MME}(Y^w_j \mid Z_j = z^w_j)\). This will be denoted as \emph{DA MME}.

	\section{Theoretical analysis}~\label{sec:analysis}
	In this section, we analyse the ability of two approaches to consensus modeling (based on the L- and D-assumptions) to output ``calibrated'' probabilities of labels. 
	
	Consider one object, assume $\mathcal{Y} = \{0,1\}$. Consider the generative process for the DA-model as described in Section~\ref{sec:latent_distribution_concept}, where \(z \sim \Ber(q)\), and \(q\) is an unknown object-specific parameter.
	Assume $n$ workers provide labels for the object independently, and each of them makes a mistake with a known probability \(1-a\), which is the same for all workers.\\
	\noindent\textbf{L-assumption approach.} According to the traditional approach, the unknown parameter \(q\) is estimated by the posterior probability of label \(1\) for the object:
	\(\hat q^{\rm LA}(n_1) := \frac{r a^{n_1}(1 -a)^{n - n_1}}{r a^{n_1} (1 -a)^{n - n_1} + (1 - r)(1 -a)^{n_1} a^{n - n_1}},\)
	where $r$ is the prior probability of label \(1\) (common for all objects) and $n_1$ is the number of ones among \(n\) noisy labels.
	
	\noindent\textbf{Proposition~1.} \emph{ The value of \( \hat q^{\rm LA}\) is neither unbiased, nor consistent estimate of \(q\).}\\
	\emph{Proof.} 
	The generative process based on the D-assumption defines the following distribution of the number of ones \(N_1\):
	\(\Pro^{\rm DA}(N_1= n_1) = {n \choose n_1} [q a + (1 - q)(1 -a)]^{n_1}[q(1 -a) +(1 - q)a]^{n - n_1}\).
	Consider the expectation of \(\hat q^{\rm LA}\) with respect to this distribution: \(\Exp (\hat q^{\rm LA}) = \sum_{n_1 = 0}^n \hat q^{\rm LA}(n_1) \Pro^{\rm DA}(N_1= n_1) \).
	One can see that this is a biased estimate for \(q\) (see the details in Appendix~\ref{appendix:theory}).
	
	Now we check the convergence of \(\hat q^{\rm LA}\) to \(q\) as \(n\) tends to infinity. By the law of large numbers, the fraction of ones among noisy labels converges:
	\(\plim_{n\to\infty} \frac{n_1}{n} = qa + (1-q)(1 -a)\). Assume that workers are not malicious, i.e. \(a > 0.5\). Note that \(\hat q^{\rm LA} =\frac{1}{1  + \frac{1 - r}{r}\left(\frac{a}{1 - a}\right)^{n\left(1 - 2n_1/n\right)}}\),
	therefore \(\hat q^{\rm LA}\) converges to different values depending on \(q\) (see the details in Appendix~\ref{appendix:theory}).
	Thus, \(\hat q^{\rm LA}\) is not a consistent estimate for \(q\) unless  the distribution \(\Ber(q)\) is degenerate or \(r = q = 0.5\). \hfill\(\square\)
	
	\noindent\textbf{D-assumption approach. }
	According to the novel approach based on the D-assumption, $N_1$ is a binomial random variable with parameters $(n, f(q))$, where $f(q):=qa+(1-q)(1-a)$ is the probability that an observed noisy label equals $1$. The estimate of parameter \(q\) is the value \(\hat q\) providing the maximum for the log likelihood of the observed noisy labels:
	\(
	\hat q^{\rm DA}(n_1) := \arg\max_{\hat q} [n_1 \log(f(\hat{q})) + (n - n_1)\log(1-f(\hat q))].
	\)
	
	\noindent\textbf{Proposition~2.} \emph{ The value of \(\hat q^{\rm DA}\) is an unbiased and consistent estimate of $q$.}\\
	\emph{Proof.} Note that \(\hat q^{\rm DA} := f^{-1}(\hat s)\), where \(\hat s :=\frac{n_1}{n}\). Therefore, we have
	\( \Exp \hat q^{\rm DA} = \Exp f^{-1}(\hat s) = f^{-1}(\Exp \hat s) =  f^{-1}(\frac{1}{n}\Exp N_1) = q.
	\)
	The second equality uses linearity of $f^{-1}$. Note that $\hat s=n_1/n$ is a consistent estimate of the success probability $f(q)$ of binomial $N_1$. Therefore, \(\hat q^{\rm DA} = f^{-1}(\hat s)\) is a consistent estimate of \(f^{-1}(f(q))=q\).\hfill\(\square\)
	
	\section{Details of implementation}
	In this section we describe our implementation for the six consensus models described in Sections~3.1--3.3. 
	Our implementation allows us to directly maximise the log-likelihood of observed noisy labels instead of using EM, which optimises the lower bound for the log-likelihood.
	
	For LA-models the log-likelihood \eqref{eq:target_ll} is a function of parameters \(\{e^w\}_{w \in W},\{d_j\}_{j \in J}\), and \( p\). To maximise the log-likelihood we use a standard conjugate gradient descent algorithm. And to compute gradients at each iteration we use a proprietary library for the automatic differentiation, which is an implementation of the operator overloading approach \citep[see, e.g.,][]{bartholomewbiggs/etal:2000}. 
	Finally, using the maximum likelihood estimates for the parameters, we compute the consensus output for each object \(j\) as the posterior probabilities for all \(z \in \mathcal{Y}\):
	\(\Pro^{\rm LA}(Z_j = z) \propto \Pro(z\mid p) \prod_{w\in W_j} \Pro(y_j^w \mid z, e^w, d_j)\)
	
	For DA-models the log-likelihood \eqref{eq:target_ld} is a function of parameters \(\{e^w\}_{w \in W},\{d_j\}_{j \in J}\), and \( \{q_j\}_{j \in J}\) and the same approach is used to maximize it. The consensus output for each object \(j\) is the estimated distribution of the object's subjective labels, i.e. for all \(z \in \mathcal{Y}\):
	\(\Pro^{\rm DA}(Z_j = z) = \Pro(z\mid q_j)\)
	
	The output of any consensus model depends on the initial values for parameters. However, it will not influence our comparisons of LA- and DA-versions of models: the same initial values are used for common parameters in each pair of models. For the pair of DS models, confusion matrices are initialised as follows:  for each object \(j\) we first compute the vector of relative frequencies \(q^{\rm RFE}_j\) of noisy labels for the object, then for each worker \(w\) we compute a matrix of counts by considering each label of the worker and adding the corresponding vector \(q^{\rm RFE}_j\) to the column \(y^w_j\) of the worker's matrix of counts, finally we normalise each row in the worker's matrix of counts to make it sum up to 1.
	Parameter \(p\) for LA DS is initialised by the frequency of each label value among all noisy labels.
	For the pair of GLAD models, workers' parameters are initialised by 1 and objects' parameters are initialised by \(\exp(1)\) as suggested in the original paper \citep{whitehill/etal:2009}.
	For the pair of MME models, workers' confusion matrices are initialised by computing the matrices of counts for workers for each worker (in the same way as for DS models) and then taking the logarithm for each element of the matrices\footnote{To avoid zeros in the matrices we add 1 to each element in the matrices of counts for workers and objects.}, and objects' parameters are initialised in a similar way: for each worker \(w\) we compute \(r^{\rm RFE}_w\) the vector of relative frequencies of noisy labels from this worker, then for each object \(j\) we compute its matrix of counts by considering each noisy label for this object and adding the corresponding vector \(r^{\rm RFE}_w\) to the row \(y^w_j\) of the matrix of counts, finally we take the logarithm for each elements in the matrix\footnote{Such nontrivial initialisation for confusion matrices in DS and MME models leads to better results comparing to trivial uniform initialisation for all matrix elements.}. Parameters of objects' latent distributions, \(\{q_j\}_{j \in J}\), in all DA-models are initialised by the frequency of noisy labels for each object \(j\).
	
	\begin{table*}[tb]
		\centering
		\begin{small}
			\begin{tabular}{lccccccr}
				\toprule
				& & \multicolumn{2}{c}{Dawid \& Skene} & \multicolumn{2}{c}{GLAD} & \multicolumn{2}{c}{Minimax entropy}\\
				\cmidrule(lr){3-4}
				\cmidrule(lr){5-6}
				\cmidrule(lr){7-8}
				Data set (number of classes) & RFE& LA & DA  &  LA & DA & LA & DA \\
				\midrule
				%
				Duchenne smiles (\(K = 2\)) & 72.08 & +4.65 & +4.03 & +3.4 & \textbf{+5.28} & -31.82 & +0.88\\
				Web search (\(K = 5\)) & 73.03 & \textbf{+12.69} & +6.66 & +8.2 & +7.3 & +3.04 & +10.88\\
				TREC (\(K = 3\)) & 45.48 & \textbf{+6.43} & +3.47 & -0.78 & -0.1 & +0.72 & +1.51\\
				Textual entailment (\(K = 2\)) & 89.92 & \textbf{+2.7} & +2.33 & +3.2 & +2.08 & -5.55 & -2.42\\
				Temporal ordering (\(K = 2\)) & 93.55 & +0.61 & \textbf{+0.82} & +0.39 & \textbf{+0.82} & -2.21 & -1.56\\
				Adult content (\(K = 4\)) & 76.04 & -0.36 & +1.44 & -0.36 & -0.96 & -1.56 & \textbf{+2.34}\\
				Price(\(K = 7\)) & 32.5 & -1.25 & +1.25 & 0.0 & 0.0 & -2.5 & \textbf{+2.5}\\
				\bottomrule
			\end{tabular}
		\end{small}
		\caption{Accuracy (in \%) of various models across real data sets. Results for the six nontrivial models are shown relative to the RFE result.  The best results for each data set are shown in bold.}\label{tab:acc}
		\vspace{-1mm}
	\end{table*}
	
	\begin{table*}[tb]
		\centering
		\begin{small}
			\begin{tabular}{lccccccr}
				\toprule
				& & \multicolumn{2}{c}{Dawid \& Skene} & \multicolumn{2}{c}{GLAD} & \multicolumn{2}{c}{Minimax entropy}\\
				\cmidrule(lr){3-4}
				\cmidrule(lr){5-6}
				\cmidrule(lr){7-8}
				Data set (number of classes) & RFE & LA & DA  &  LA & DA & LA & DA \\
				\midrule
				%
				Duchenne smiles (\(K = 2\)) & \(\infty\) & 2.467 & 0.821 & 2.074 & \textbf{0.753} & 3.464 & 0.817\\
				Web search (\(K = 5\)) & \(\infty\) & \textbf{0.247} & 0.364 & 0.408 & 0.468 & 0.747 & 0.331\\
				TREC (\(K = 3\)) & \(\infty\) & 1.403 & 1.143 & 1.28 & \textbf{1.116} & 1.779 & 1.196\\
				Textual entailment (\(K = 2\)) & 0.509 & 0.554 & 0.363 & \textbf{0.308} & 0.377 & 0.9 & 0.454\\
				Temporal ordering (\(K = 2\)) & \(\infty\) & 0.841 & 0.26 & 0.331 & \textbf{0.252} & 0.817 & 0.286\\
				Adult content (\(K = 4\)) & \(\infty\) & 1.345 & 0.507 & 0.875 & 0.535 & 1.385 & \textbf{0.501}\\
				Price (\(K = 7\)) & \(\infty\) & 80.365 & 3.128 & 16.357 & \textbf{0.917} & 42.651 & 1.675\\
				\bottomrule
			\end{tabular}
		\end{small}
		\caption{Log loss results across real data sets. The best results for each data set are shown in bold.}
		\label{tab:nll}
		\vspace{-5mm}
	\end{table*}

	\section{Experiments}\label{sec:empirical}
	In this section we empirically show that:
	(1) the novel D-assumption is more realistic than the L-one, and, as a result, DA-models produce more accurate consensus labels;
	(2) models designed for the D-assumption are capable for recovering latent distributions of labels. A straightforward way to check the first proposition, is to compare performance of an LA-model against the corresponding DA-counterpart.
	However, as distributions of labels for real objects are unknown, we use synthetic data sets to check the second statement.
	
	Consider multiclass labels \(\{1,\ldots, K\}\). For each object \(j\), a model M produces a consensus label \(\hat q^{\rm M}_j\), that is a vector of length \(K\) and its \(k\)-th element \(\hat q^M_j[k]\) is the probability for the label value \(k\). Note that all models, LA- and DA-, produce probabilistic labels, but, as we explained in the previous sections, the intuition behind the LA-model probabilistic output is ``confidence'' in each label value, whereas for the DA-one it is estimated probability of labels.
	
	\paragraph{Relative frequency estimator (RFE)} Given a set of noisy labels for an object \(j\), a standard baseline for consensus modeling is to estimate probabilities for all \(z \in \mathcal{Y}\) as
	\(\Pro^{\rm RFE}(Z_j = z) := \frac{\left|\{w\in W_j: y^w_j = z\}\}\right|}{\left|W_j\right|}\), where \(W_j\) is the set of indices of workers that label the object \(j\).
	
	For each data set, the \emph{training set} is a set of objects with multiple noisy labels for each object, and the \emph{test set} is a subset of those objects with a known ground truth label for each of them. The quality of consensus labels produced from the training set is evaluated over the test set. Let \(T\) be a set of indices of the test set objects, and, for an object \(i \in T\), \(t_i\) be its ground truth label. We use the following two metrics to evaluate performance of a model on the test set:\\
	\noindent\emph{Accuracy. }
	For each object \(j\) in a training set, a model M produces a probabilistic label \(\hat q^{\rm M}\), and an estimated label for this object is \(\hat t^{\rm M}_j := \arg \max_k \hat q^{\rm M}_j[k]\).\footnote{If the maximum is attained at several label values, we randomly choose one of them. In our experiments, RFE accuracy is averaged over 10 runs to reduce the effect of random tie-breaking. There was no need to average results of the nontrivial models as no tie was observed.}  Accuracy over a test set is defined as 
	\(
	{\rm acc} := \frac{1}{\left|T\right|} \sum_{j \in T} \mathds{I}(t_j = \hat t^{\rm M}_j),
	\)
	where \(\mathds{I}(X)\) is the indicator function for \(X\).\\
	\noindent\emph{Log loss. } 
	Given consensus labels \(\hat q_j^{\rm M}\) based on a model M, the log loss over a test set is the mean negative log likelihood of the test set labels,
	computed as
	\({\rm logloss} := -\frac{1}{\left|T\right|} \sum_{j \in T} \log_{K}{\left(q^{\rm M}_j[t_j]\right)}\).
	\footnote{For this definition, log loss for non-informative probabilistic labels, consisting of probabilities \(\frac{1}{K}\), is 1.}
	

	\begin{figure*}[tb]
		\centering
		\includegraphics[width=0.7\textwidth]{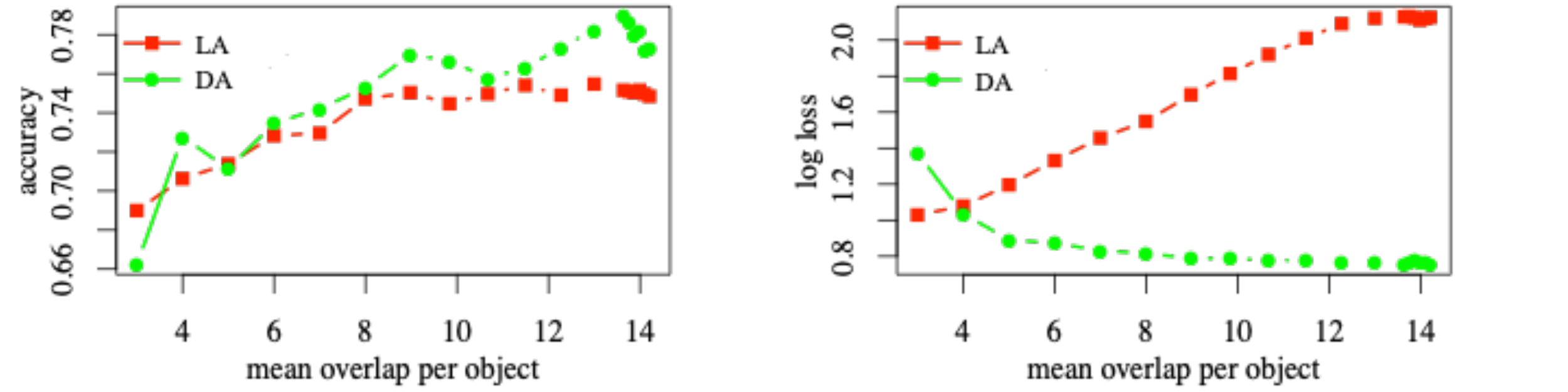}
		\caption{Performance of the LA GLAD (red line with squares) and DA GLAD (green line with dots) on the Duchenne smiles data set. The horizontal axis is for the mean number of labels per object. }
		\label{fig:overlap}
		\vspace{-5mm}
	\end{figure*}

	\noindent{\textbf{Real-world data.} } First, compare the seven models across the following data sets: Duchenne smiles \citep{whitehill/etal:2009}, Web search \citep{zhou/etal:2015}, TREC \citep{buckley/etal:2010}, Textual Entailment \citep{snow/etal:2008}, Temporal Ordering \citep{snow/etal:2008}, Adult content \citep{ipeirotis/provost/wang:2010}, Price \citep{liu/ihler/steyvers:2013}. For each data set, ground truth labels for a subset of objects are provided by the domain experts. (See Appendix~\ref{appendix:data} for the tasks description and summary statistics of the data sets.) 
	Table~\ref{tab:acc} shows accuracy results. For a data set, accuracy of each nontrivial model is shown as the difference between the result for the RFE baseline (given in the second column) and the model result. As a summary, among 21 comparisons of LA- and DA-models, the DA-counterpart is better in 13 out of 21 times. Besides this, the novel approach generates the best model for 4 out of 7 data sets.
	Table~\ref{tab:nll} shows the log loss of the seven models \footnote{To avoid infinite losses for RFE we applied add-one smoothing to \(q_j^{\rm RFE}\).} across the six data set. Remarkably, for the log loss, among 21 comparisons of LA- and DA-models, the DA-counterpart is better in 18 out of 21 times. And for 5 out of 7 data sets, the novel approach generates the best model.
	Let us highlight some interesting observations from Tables~\ref{tab:acc} and \ref{tab:nll}. 
	First, consider the Textual entailment data set: according to the log loss, performance of the RFE baseline is infinitely bad\footnote{This happens if none of noisy labels for an object is equal to the ground truth.} for all the data sets except the Textual entailment, which may indicate that the task was easy, in the sense that the traditional L-assumption is likely to be true, and the DA-models suffer due to their greater flexibility. Secondly, consider the TREC data set, where according to the log loss, the quality of probabilities obtained by any model, LA or DA, is worse than that for non-informative uniform probabilities. We checked, that the mean number of labels per object for this data set is \(4.6\) which may be not enough to get realistic distributions over three classes. And similarly we explain the results for Web search data set with the mean number of labels per object is \(5.8\).

	\noindent{\textbf{The number of labels per object.} } To analyse how the number of labels per object affects the performance of the LA- and DA-models, we conducted the following experiment. As an example, we take the Duchenne smiles data set and two models, LA GLAD and DA GLAD. Figure~\ref{fig:overlap} shows accuracy and the log loss for the Duchenne smiles data as the mean number of labels per object grows from \(3\) to \(14.2\). For each integer \(n \ \in [3;20]\), we sampled a part of the data set, such that the number of labels per object not greater than \(n\), it  was done as following: given a number of noisy labels for an object, we randomly draw \(n\) of those labels (if the number of noisy labels for the object was \(\leq n\), all the noisy labels were taken). For the sample of the data set, we obtained consensus labels using the two models and evaluate their performance as before. 
	Results shown in Figure~\ref{fig:overlap} are averaged over \(10\) simulations. Indeed, when the number of noisy labels is \(3\), the DA-model is worse than the LA-one according to both performance metrics.
	However, as the number of labels per object increases, the performance of the DA-model notably exceeds that of the LA-model at almost all overlap values. Especially, for the log loss, we can see that for the DA-model, the bigger the overlap of noisy labels per object, the better the probabilistic labels. Whereas, for the LA-model, when the number of labels per object grows, aggregated probabilistic labels converge to deterministic values and the log loss for the test set becomes worse.
	
	\begin{figure}
		\begin{minipage}{.4\textwidth}
			\centering
			
				\begin{tabular}{cc}
				\toprule
				Model  & MSE\\
				\midrule
				LA & \(0.071 \pm 0.002\)\\
				DA &  \(0.010 \pm 0.001\) \\
				\bottomrule
			\end{tabular}

			\captionof{table}{ Mean squared error between true and estimated probabilities. Rows are for approaches to consensus modelling. The data generating process is based on the D-assumption.}\label{tab:mse}
		\end{minipage}
		\hfill
		\begin{minipage}{.58\textwidth}
			\captionsetup[subfigure]{position=b}
			\centering
			\begin{subfigure}[b]{.4\textwidth}
				\centering
				\resizebox{\textwidth}{!}{\includegraphics[width=\textwidth]{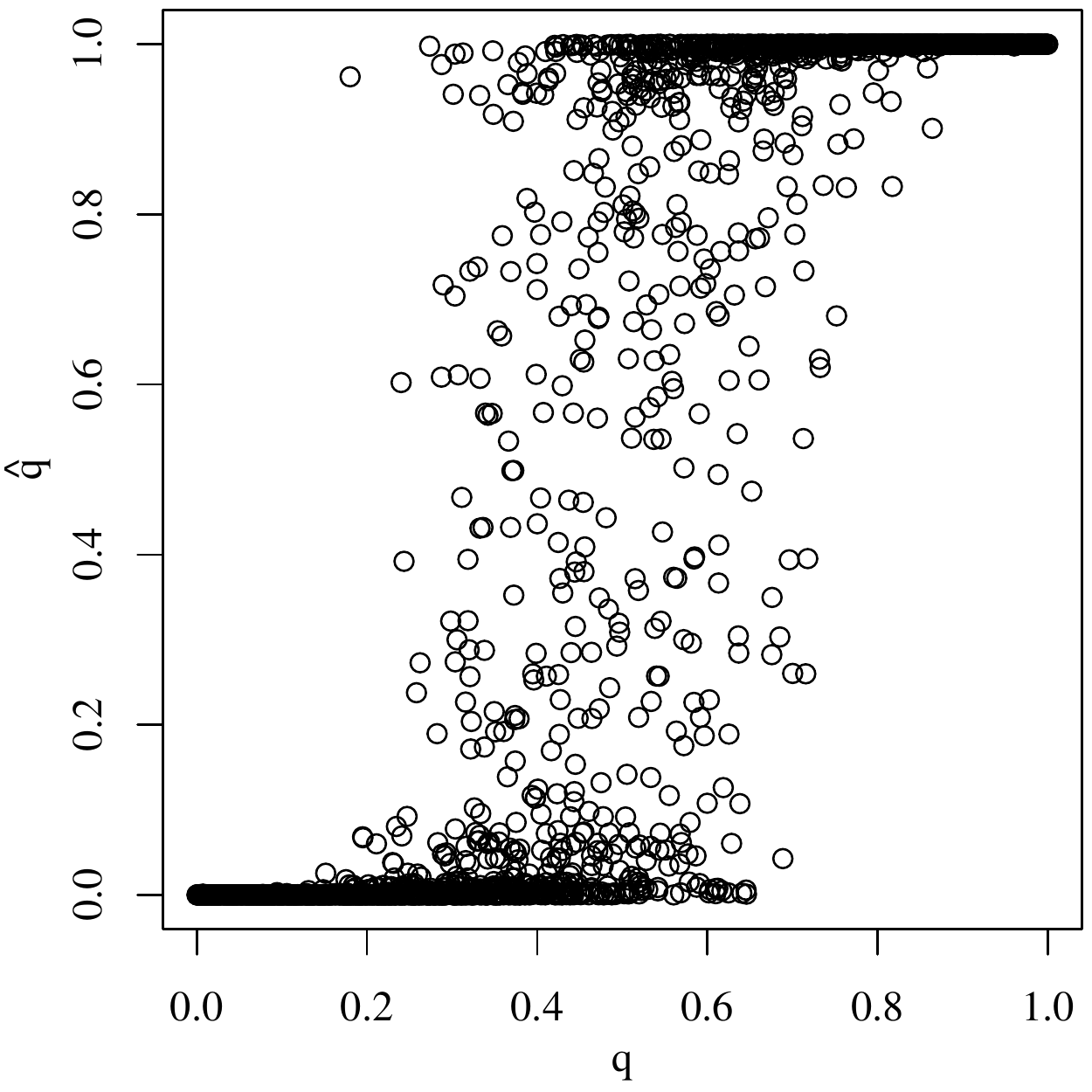}}
				\caption{LA-model.}
				\label{fig:calib_ll_ll}
			\end{subfigure}\hfill
			\begin{subfigure}[b]{.4\textwidth}
				\centering
				\resizebox{\textwidth}{!}{\includegraphics[width=\textwidth]{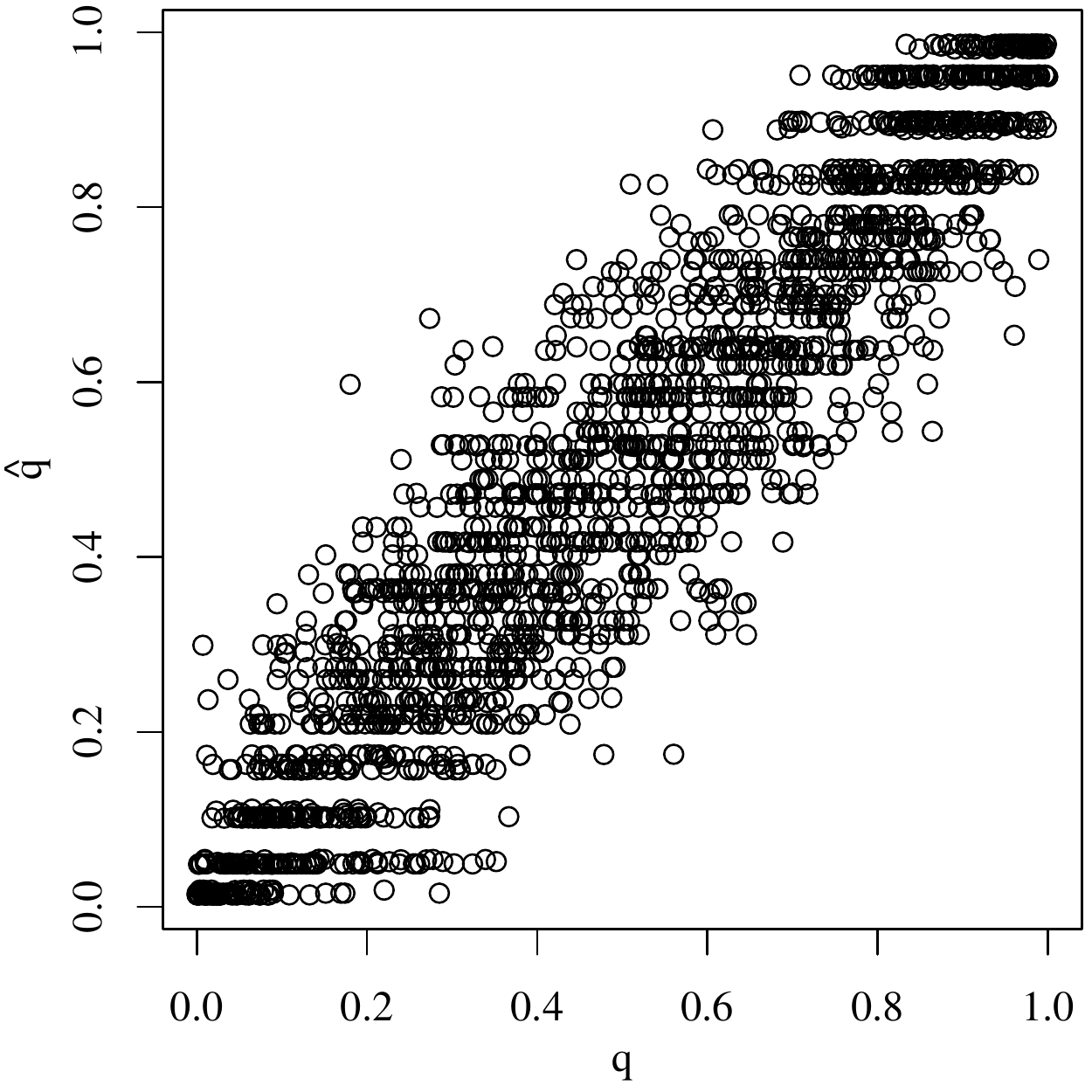}}
				\caption{DA-model.}
				\label{fig:calib_ll_ll}
			\end{subfigure}
			\caption{Calibration plots for two approaches to consensus modelling. The horizontal axis is for the true probabilities and the vertical one is for the estimated probabilities.}
			\centering
			\label{fig:calibration}
		\end{minipage}
		\vspace{-7mm}
	\end{figure}
	
	\noindent{\textbf{Synthetic data.} } Inspired by the analysis in Section~\ref{sec:analysis}, next we empirically examine the ability of the two approaches for consensus modelling to recover the underlying probabilities for labels.
	Data for this experiment is generated similarly to \citep{whitehill/etal:2009}: we used \(N = 2000\) objects and 20 workers, each worker labels each object; for each object we randomly generate its latent parameter \(q_j \sim \Unif[0,1]\), which is the probability of label \(1\); to produce noisy labels we generate the following parameters for GLAD \(e^w \sim \Unif[0,4]\) and \(\log(d_j) \sim \Unif[0,3]\). 
	Using the parameters, a data set is generated according to the DA GLAD. Then, for the data set, we used our implementations for the LA GLAD and DA GLAD to obtain estimates \(\hat q_j\). Table~\ref{tab:mse} shows the mean squared error, \(\frac{1}{N}\sum_{j=1}^n(\hat q_j - q_j)^2\), averaged across 10 simulations. This confirms that DA-model is able to accurately estimate the underlying true probabilities.
	Figure~\ref{fig:calibration} shows calibrations plot for estimated probabilities obtained using the two approaches to consensus modelling. It demonstrates that probabilities estimated by the DA-model are empirically well calibrated, whereas the LA-model is biased towards degenerate distributions.
	
	\section{Conclusion}\label{sec:conclusion}
	We have proposed the novel D-assumption for consensus modelling and a way of constructing novel DA-models under the assumption. 
	It has been shown that the DA-models are able to obtain proper estimates for true probabilities in contrast to those for the established LA-models. Our experiments confirm that the DA-models often perform better for real data sets, implying that, 
	the D-assumption is more realistic than the L-one.
	However, there is no universal model which is the best for every data set (even among the state-of-the-art models). Thus, it
	always makes sense to produce golden labels (by professional judges) at least for a small part of any data set to evaluate
	all existing models and check which one serves the best for the data at hand. In that sense, our novel D-assumption approach is
	complimentary to the family of existing consensus models and allows to twice increase their number via considering our alternative noisy label generation process to be implemented for each one of them.
	
	\bibliography{bibliography}
	\bibliographystyle{plain}
	\begin{appendices}
	\section{Theoretical analysis for the latent label assumption}\label{appendix:theory}
	Remind, that we consider one object whose ``true'' label \(z\sim\Ber(q)\), where \(q\) is an unknown object-specific parameter. Given $n$ noisy labels for this object, such that each noisy label is correct with a known probability \(a\), the goal is to estimate the parameter \(q\).\\
	
	According to the LA-model, the unknown parameter \(q\) is estimated by the posterior probability of label \(1\) for the object:
	\begin{align*}
		&\hat q^{\rm LA}(n_1) := \frac{r a^{n_1}(1 -a)^{n - n_1}}{r a^{n_1} (1 -a)^{n - n_1} + (1 - r)(1 -a)^{n_1} a^{n - n_1}},
	\end{align*}
	where $r$ is the prior probability of label \(1\) (common for all objects) and $n_1$ is the value of the random variable \(N_1\), which is the number of ones among \(n\) noisy labels.
	\begin{prop}
		The posterior probability of label \(1\) obtained under the L-assumption is neither unbiased, nor consistent estimate of the true probability $q$ underlying the generative process for the DA-model.
	\end{prop}
	\begin{prf}
		The posterior probability of label \(1\) for the object is
		\begin{align*}
			&\hat q^{\rm LA}(n_1) := \frac{r a^{n_1}(1 -a)^{n - n_1}}{r a^{n_1} (1 -a)^{n - n_1} + (1 - r)(1 -a)^{n_1} a^{n - n_1}},
		\end{align*}
		where $r$ is the prior probability of label \(1\) (common for all objects) and $n_1$ is the value of the random variable \(N_1\), which is the number of ones among \(n\) noisy labels.
		
		The generative process based on the D-assumption defines the following distribution over the number of ones \(n_1\):
		\begin{multline*}
			\Pro^{\rm DA}(N_1= n_1 \mid N = n) := {n \choose n_1} [q a + (1 - q)(1 -a)]^{n_1}[q(1 -a) +(1 - q)a]^{n - n_1}.
		\end{multline*}
		Consider the expectation of \(\hat q^{\rm LA}\) with respect to this distribution:
		\begin{multline}\label{eq:exp_p_old_under_novel}
			\Exp (\hat q^{\rm LA}) := \sum_{n_1 = 0}^n \hat q^{\rm LA}(n_1) \Pro^{\rm DA}(N_1= n_1 \mid N = n )\\
			= \sum_{n_1 = 0}^n \Bigg[\frac{{n \choose n_1}[q a + (1 - q)(1 -a)]^{n_1}[q(1 -a) +(1 - q)a]^{n - n_1}}{r a^{n_1} (1 -a)^{n - n_1} + (1 - r)(1 -a)^{n_1} a^{n - n_1}} r a^{n_1}(1 -a)^{n - n_1}\Bigg]
		\end{multline}
		
		To demonstrate that this is a biased estimate of the true parameter \(q\), evaluate this expression as a function of \(q\) for different values of \(a\) and the uniform prior \(r = 0.5\). Figure~\ref{fig:traditional_estimate}, showing the dependence of the expected value for \(\hat q^{\rm LA}\) from \(q\), confirms that \(\hat q^{\rm LA}\) is a biased estimate for \(q\).
		
		\begin{figure*}[tb]
			\begin{center}
				\includegraphics[width=0.8\textwidth]{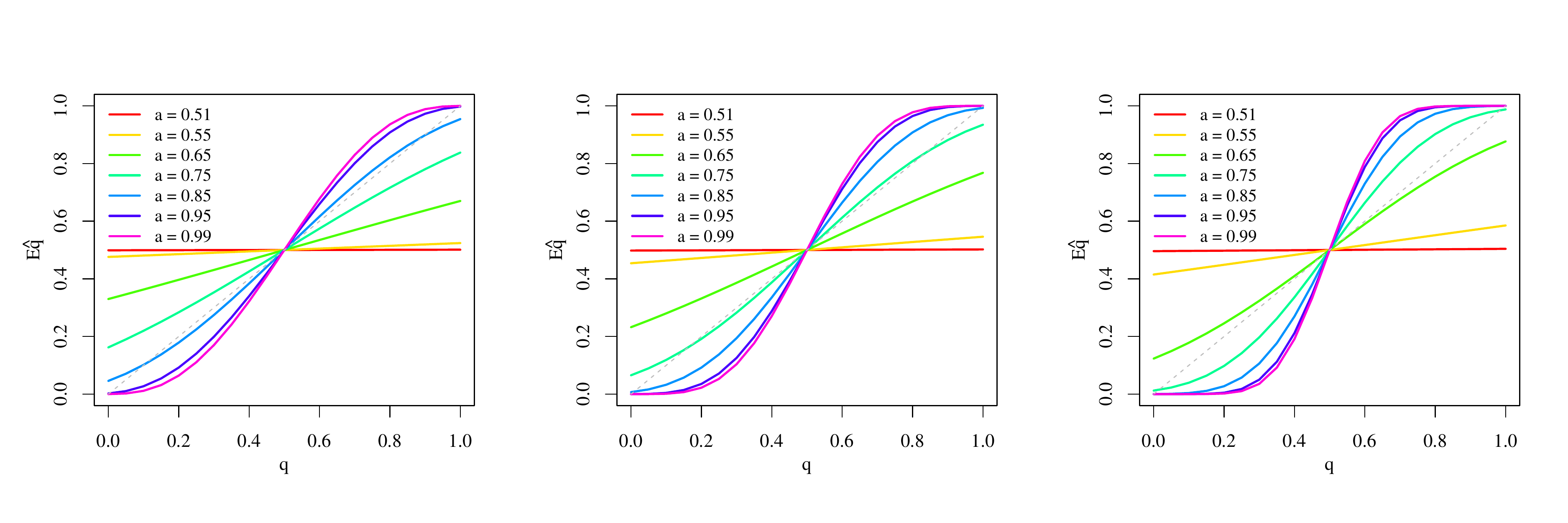}
				\caption{The expected values for \(\hat q^{\rm LA}\) given by \eqref{eq:exp_p_old_under_novel} as a function of \(q\) for different number of noisy labels \(n\), different values of \(a\), and the uniform prior \(r = 0.5\). The left plot is for \(n = 5\), the middle one is for \(n = 10\), and the right one is for \(n = 20\). Different values of \(a\) are shown by colours.}
				\label{fig:traditional_estimate}
			\end{center}
		\end{figure*}
		Now we check whether \(\hat q^{\rm LA}\) converges to \(q\) as the number of noisy labels \(n\) tends to infinity. By the law of large numbers, the fraction of ones among noisy labels converges:
		\begin{equation*}
			\plim_{n\to\infty} \frac{n_1}{n} = qa + (1-q)(1 -a).
		\end{equation*}
		
		Assume that workers are not malicious, i.e. \(a > 0.5\). Note that
		\begin{align*}
			\hat q^{\rm LA} =\frac{1}{1  + \frac{1 - r}{r}\left(\frac{a}{1 - a}\right)^{n\left(1 - 2n_1/n\right)}},
		\end{align*}
		therefore \(\hat q^{\rm LA}\) converges to different values depending on \(q\):
		\begin{multline}
			\plim_{n\to \infty}\hat q^{\rm LA} = \plim_{n\to \infty}\frac{1}{1  + \frac{1 - r}{r}\left[\left(\frac{a}{1 - a}\right)^{\left(1 - 2(qa + (1-q)(1 -a))\right)}\right]^n} = \left\{
			\begin{array}{ll}
				0, & {\rm if}~ q < \frac{2a-1}{4a-2}=0.5; \\
				r, & {\rm if}~ q =0.5; \\
				1, & {\rm if}~ q > 0.5.
			\end{array}
			\right.
		\end{multline}
		Thus, \(\hat q^{\rm LA}\) is not a consistent estimate for \(q\) unless  the distribution \(\Ber(q)\) is degenerate or \(r = q = 0.5\).
	\end{prf}
	
	\section{Data sets details}\label{appendix:data}
	We use the following public data sets for our empirical studies:
	\begin{itemize}
		\item \textbf{Duchenne smiles} \cite{whitehill/etal:2009}.
		The task is to classify images into two categories -- a Duchenne smile (``enjoyment'' smile) and  a non-Duchenne (``social'' smile). 
		\item \textbf{Web search} \citep{zhou/etal:2015} and \textbf{TREC} \citep{buckley/etal:2010}.
		The task is to rate query-URL pairs. For a given query-URL pair, a worker is asked to rate how relevant is the URL to the search query. For the Web search data, the rating scale has 5 levels: perfect, excellent, good, fair, or bad. For the TREC data, the rating scale is ternary: highly relevant, relevant, and non-relevant.
		\item \textbf{Recognising Textual Entailment} \citep{snow/etal:2008}. For this task each object contains two statement, and a worker judges whether one statement implies another.
		\item \textbf{Temporal Ordering} \citep{snow/etal:2008}. Each object describes two events and the task is to judge whether one event follows another.
		\item \textbf{Adult content} \citep{ipeirotis/provost/wang:2010}. The task is to classify web pages into four categories depending on the presence of adult content on them.
		\item \textbf{Price} \citep{liu/ihler/steyvers:2013}. The task is to estimate prices of household items choosing one out of seven adjacent bins corresponding to different ranges of price.
	\end{itemize}
	
	\begin{table*}[tb]
		\caption{Summary statistics for real data sets.}
		\label{tab:data_statistics}
		\centering
		\begin{tabular}{lcccccccr}
			\toprule
			Data set & \# Cl. & \# Obj. & \# Workers & \# Samples & Lab. per obj. & \multicolumn{2}{c}{Lab. per worker}& Gr. truth\\
			\cmidrule(lr){7-8}
			&   &   &  & & mean & mean & med. & objects\\
			\midrule
			Duchenne smiles & 2 & 2134 & 64 & 30319 & 14.2 & 473.7 & 109 & 159\\
			Web search & 5 & 2665 & 177 & 15567 & 5.8 & 87.9 & 19 & 2653\\
			TREC & 3 & 20026 & 762 & 91783 & 4.6 & 120.5 & 18 & 3275\\
			Textual entailment & 2 & 800 & 164 & 8000 & 10.0 & 48.8 & 20 & 800\\
			Temporal ordering & 2 & 462 & 76 & 4620 & 10.0 & 60.8 & 20 & 462\\
			Adult content & 4 & 11040 & 825 & 92721 & 8.4 & 112.4 & 18 & 1517\\
			Price & 7 & 80 & 155 & 12400 & 155.0 & 80.0 & 80 & 80\\
			\bottomrule
		\end{tabular}
	\end{table*}
	Table\ref{tab:data_statistics} provides summary statistics for all the data sets. The table includes the following columns: the size of the label set, the number of objects, the number of crowdsourcing workers, the total number of noisy labels from those workers -- the size of the training set, the mean number of labels per object, the mean and the median number of labels per worker, and also the number of objects with known ground truth labels -- the size of the test set.
	\end{appendices}
	
\end{document}